\documentclass[prb,amsmath,amssymb,showpacs,footnote,nofootinbib,twocolumn]{revtex4}

\usepackage{graphics}
\usepackage{epsfig}
\usepackage{graphicx}
\usepackage{dcolumn}
\usepackage{bm}
\usepackage[english]{babel}

\def\Ref#1{(\ref{#1})}

\begin{document}

\title{Shear Induced Structural Ordering of a Model Metallic Glass}

\author{Anatolii V. Mokshin$^{1,2}$}
\author{Jean-Louis Barrat$^1$}

\address{$^1$Universit\'e de Lyon; Univ. Lyon
I,  Laboratoire de Physique de la Mati\`ere Condens\'ee et des
Nanostructures; CNRS, UMR 5586, 43 Bvd. du 11 Nov. 1918, 69622
Villeurbanne Cedex, France}

\address{$^2$Department of Physics, Kazan State University, Kremlyovskaya 18,
420008 Kazan, Russia}

\date{\today}

\begin{abstract}
We report results of non-equilibrium molecular dynamics
simulations of a  one-component glassy system under the influence
of a shear flow, with the aim of investigating shear induced
ordering of this system.  In spite of the very low temperature,
the system transforms into a strained crystalline state through
well defined nucleation events. Various characteristics of the
observed ordering at  different strain rates and temperatures are
discussed. We also define and discuss the  transition rates.
\end{abstract}

\maketitle

Nucleation, clustering and ordering processes in supercooled melts
are  a subject of considerable interest, both in soft matter and
in hard materials such as metallic alloys or molecular glass
formers. Crystal ordering is generally thought to proceed through
nucleation events, which are described within the framework of
classical nucleation theory (CNT)
\cite{Farkas,Becker,Zeldovich,Turnbull}. In this theory, the
nucleation is associated with the local crossing of a free energy
barrier under the influence of thermal noise. The situation
becomes  more intriguing if, instead of a supercooled liquid,   we
consider a glassy system at very deep supercooling
\cite{Schmelzer}. The kinetic slowing down and high viscosity of a
glass make the structural rearrangements, including those required
for crystal nucleation, so slow that the corresponding transition
is practically unobservable on the experimental time scales.
Obviously, the devitrification can be accelerated through
reheating of the glass. An interesting alternative, which is in
fact commonly used in soft matter systems, is to utilize shear as
an ordering tool rather than temperature. Crystal-crystal or
amorphous crystal transformations are also known to take place in
processes such as grinding and milling of molecular materials,
which in general involve a complex combination of mechanical
stresses and temperature variations \cite{willart}.

Recent numerical studies provide the evidence of crystallization
induced by  external shear in an amorphous, single-component
Lennard-Jones system \cite{Duff,Mokshin/Barrat} However, such
single-component Lennard-Jones systems have a very weak glass
forming ability and the transition to a crystalline state at low
temperature proceeds through a spinodal, rather than nucleation,
route \cite{Trudu,Bartell_2007}.

In this study, we consider the case of a system with a much better
glass forming ability than the Lennard-Jones system, but still
keeping the simplicity associated with one-component systems. This
system is made of particles interacting through the short ranged,
oscillating  Dzugutov potential \cite{Dzugutov_PRA,Dz_link}
\begin{eqnarray}
U(r^*)/\epsilon &=& A ~({r^*}^{-m}-B)~\mathrm{exp}\left (
\frac{c}{r^*-a} \right ) \Theta(a - r^*)\nonumber \\ &+& B ~
\mathrm{exp} \left ( \frac{d}{r^* - b} \right ) \Theta(b - r^*), \
 r^*=r/\sigma, \label{Dzugutov_pot}
\end{eqnarray}
where $\sigma$ and $\epsilon$ are the length and energy scales,
respectively, the parameters $A=5.82$, $B=1.28$, $m=16$, $a=1.87$,
$b=1.94$, $c=1.1$, $d=0.27$ are taken as described in Ref.
\cite{Dzugutov_PRA} and $\Theta(r)$ represents the Heaviside step
function. So, the potential is characterized by a minimum
$U(r_{min})= - 0.581\epsilon$ located at $r_{min}=1.13\sigma$ and
a maximum $U(r_{max})=0.46\epsilon$ at $r_{max}=1.628\sigma$. The
presence of these oscillations can be considered as a simplified
attempt to reproduce the screened Coulomb repulsion of ions in
metals. The potential is tuned in such a way that it prevents the
formation of the most common crystal phases at low pressure. At
high pressures ($P\geq 5$$\epsilon/\sigma^3$) and low enough
temperatures\cite{footnote1}, the $\mathrm{FCC}$ or $\mathrm{HCP}$
crystalline structures can be formed, although the ground state of
the system at zero temperature is a $\mathrm{BCC}$ structure. A
slow quench from a high-temperature state this model favors the
formation of quasi-crystals including Frank-Kasper structures,
whereas a faster cooling  leads to a metastable amorphous state
with a mode-coupling temperature $T_{MCT}$ estimated to be
$0.4$$\epsilon/k_B$ (see Ref. \cite{Dzugutov_PRL}). A remarkable
property is that this is one-component good glass former shares
many common features with much more complex metallic glasses. With
only one component, however,  the study of topological
reorganization, which excludes such effects as chemical ordering
or demixing, is facilitated \cite{Simdyankin}.

The system consists of $N=19~652$ particles contained in a cubic
simulation box of the volume $L^3$ with $L=28.55\sigma$. The
structural changes in such a rather large system is expected to be
relatively insensitive to the boundary conditions imposed by the
simulation cell \cite{Wedekinda1}. A set of well equilibrated
independent samples is created  at a temperature
$T=2.0$$\epsilon/k_B$. These samples are cooled at a
 rate $\Delta T/\Delta t = 0.000~97$$\epsilon/(k_B \tau)$
and the configurations corresponding to the temperatures $T=0.01$,
$0.03$ and $0.06$$\epsilon/k_B$ are stored\cite{footnote2}. To
impose the shear we create two solid walls at the ends of the
simulation box in the $y$ direction by freezing all the particles
in the $x$-$z$ plane over the range of three interparticle
distances. The top wall moves in the $x$ direction with the
instantaneous velocity $\textbf{\textit{v}}(t) =
\dot{\gamma}L_y(t) \textbf{\textit{e}}_x$ at a  constant strain
rate $\dot{\gamma}$ and  normal load  $P_{yy} =
7.62$$\epsilon/\sigma^3$; $L_y$ is the distance between the walls.
This  value of the normal pressure  is thermodynamically favorable
to FCC crystalline forms and allows one to avoid the phase of
quasi-crystals for the Dzugutov system \cite{Roth_PRE_2000}.
Constant temperature conditions are ensured by rescaling the
velocity component in the neutral $z$ direction perpendicular to
the flow $x$ and the velocity gradient $y$ direction. The
termostatting was implemented only if the temperature has changed
more than $1.5\%$ of the considered value. Other details  of the
simulation protocol are identical to those described in Ref.
\cite{Mokshin/Barrat}.

The local  order in the system can be characterized by  parameters
that probe  the  appearance of orientational order around each
atom. In this  work, we use both the local orientational
\cite{Wolde} bond-order parameter $q_l(i)$ and its global
\cite{Steinhardt} counterpart $\mathcal{Q}_l$:
\begin{subequations}
\begin{equation}
q_{l}(i) = \left ( \frac{4\pi}{2l+1} \sum_{m=-l}^{l} \left |
\frac{\sum_{j=1}^{N_b(i)} Y_{lm}(\theta_{ij},
\varphi_{ij})}{N_b(i)} \right |^2 \right )^{1/2},
\label{local_order_par}
\end{equation}
\begin{equation}
\mathcal{Q}_l = \left ( \frac{4\pi}{2l+1} \sum_{m=-l}^{l} \left |
\frac{\sum_{i=1}^{N} \sum_{j=1}^{N_b(i)} Y_{lm}(\theta_{ij},
\varphi_{ij})}{\sum_{i=1}^{N} N_b(i)} \right |^2 \right )^{1/2}.
 \\ \label{global_order_par}
\end{equation}
\end{subequations} Here $Y_{lm}(\theta_{ij}, \varphi_{ij})$ are
the spherical harmonics, $N_b(i)$ denotes the number of the
nearest ``neighbors'' of particle $i$, and $\theta_{ij}$ and
$\varphi_{ij}$ are the polar and azimuthal angles formed by
radius-vector $\textbf{r}_{ij}$ and some reference system.  The
parameter $\mathcal{Q}_l$ is  invariant under rotations. Nonzero
values of $\mathcal{Q}_l$ for the different crystalline cluster
structures including those with icosahedral symmetry appear for $l
\geq 6$ (see Ref. \cite{Steinhardt}). In this work we define
``neighbors'' as all atoms located within a sphere of  radius
$|\textbf{r}_{ij}|<1.5\sigma$ around an atom $i$. For a  fully
disordered system one has $\mathcal{Q}_6 \rightarrow 0$, whereas
the ordered  geometries are characterized by the next values of
$\mathcal{Q}_6$: $\mathcal{Q}_6^{fcc} = 0.5745$,
$\mathcal{Q}_6^{bcc} = 0.5106$, $\mathcal{Q}_6^{ics} = 0.6633$,
$\mathcal{Q}_6^{hcp} = 0.4848$ and $\mathcal{Q}_6^{sc} = 0.3536$
(see Refs. \cite{Rintoul,Wolde}).

By performing a cluster analysis for the detection of the
``solidlike''particles (for details see Refs.
\cite{Mokshin/Barrat,Wolde}), we obtain the cluster size
distribution $n_s(t)$ and its moments $s^{(m)}(t)=\sum_s s^m
n_s(t)$ as a function of time. The zeroth moment $s^{(0)}$
defines the total number of  clusters, the first moment
$s^{(1)}$ represents the total number of particles involved in an
ordered phase (solidlike particles), the ratio of the second and
first moments, $s^{(2)}/s^{(1)}$, is associated with the average
cluster size. In a finite sample we also  obtain $s^* =
\mathbf{\max}_{s,\, n_s\neq 0}[n_s]$ as the size of the largest
cluster. The average \emph{spatial} cluster size $\langle \zeta
\rangle$ can be also estimated through the distance between
maximally-separated particles within  the same cluster.

\begin{figure}[tbh]
\centerline{\psfig{figure=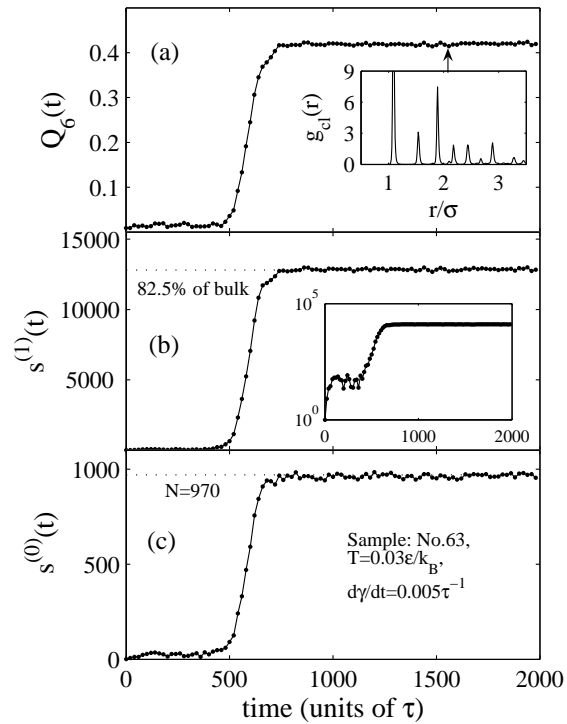,height=10cm,angle=0}}
\caption{\label{Cl_analysis} Time-dependence of different
characteristics computed for a single sheared sample with the
strain rate $\dot{\gamma}=0.005\tau^{-1}$ at the temperature
$T=0.03\epsilon/k_B$: (a) Global orientational bond-order
parameter $\mathcal{Q}_6(t)$. Inset: radial distribution of
particles formative the ordered cluster structures. It is defined
for the moment indicated by arrow. Pronounced peaks of radial
distribution appear at the distances $\sim \sigma n^{1/2}$,
 $n=1,\; 2,\; 3,\ldots$ is the number of a peak, that is
typical for FCC structures. (b) Number of solidlike particles
$s^{(1)}(t)$. Inset shows the same in logarithmic scale. (c)
Number of clusters $s^{(0)}(t)$.}
\end{figure}
\begin{figure}[tbh]
\centerline{\psfig{figure=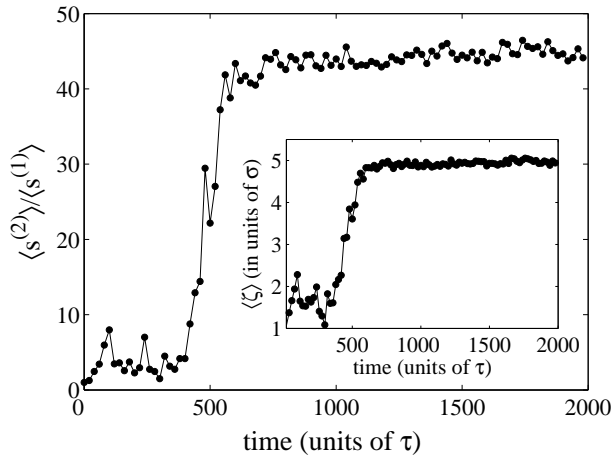,height=6.5cm,angle=0}}
\caption{\label{correl_length}  Time evolution of two average
cluster characteristics in the sheared system with the strain rate
$\dot{\gamma}=0.005\tau^{-1}$ at the temperature
$T=0.03\epsilon/k_B$. Main: Average cluster size as the ratio
between the second and first moments of the cluster size
distribution. Inset: spatial correlation length (see text).
Relatively insignificant values of both terms for the ordered
phase is consequence of the retaining of small crystalline grains
(for the range $t>600\tau$ the ratio between the largest cluster
$s^*$ presented in Fig.~\ref{largest_cl} and the average cluster
size $s^{(2)}/s^{(1)}$ is $\sim 4$ times).}
\end{figure}
\begin{figure}[tbh]
\centerline{\psfig{figure=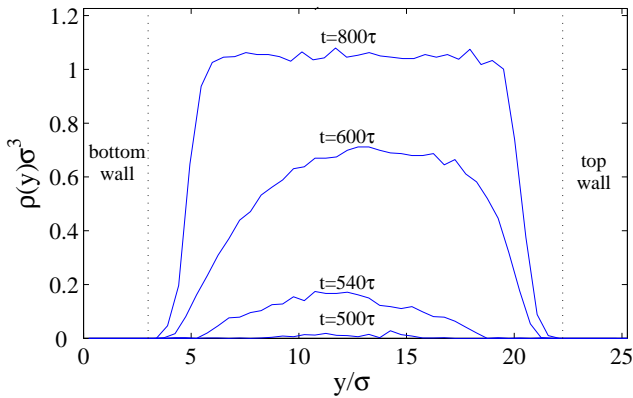,height=7cm,angle=0}}
\caption{\label{profile} (Color online) Density profiles of
solidlike particles for the instantaneous configurations at the
different times after shear initiation. The results are typical
for all studied samples. Profiles presented here are for the same
system as in the previous figure.}
\end{figure}
\begin{figure}[tbh]
\centerline{\psfig{figure=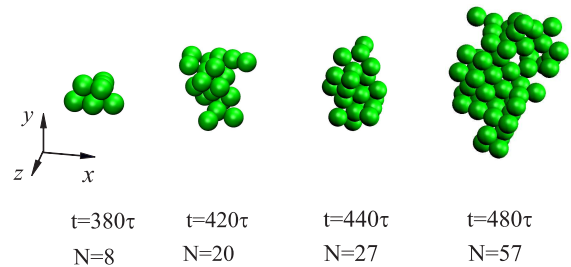,height=3.5cm,angle=0}}
\caption{\label{Cl_evol} (Color online) Largest cluster at the
different time steps prior to transition. $N$ indicates the size.
The sample is the same as for the previous figures.}
\end{figure}
\begin{figure}[tbh]
\centerline{\psfig{figure=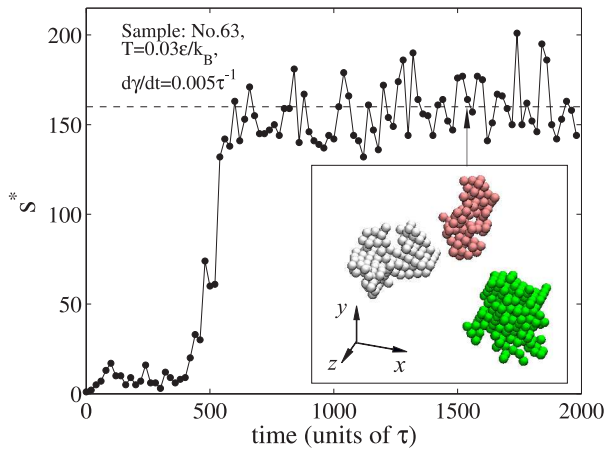,height=6cm,angle=0}}
\caption{\label{largest_cl} (Color online) Main: Size of the
largest cluster \textit{versus} time. Broken line corresponds to
the value $160$. The study sample is still the same as for the
previous figures. Inset: Three largest clusters at the moment
indicated by arrow. The sizes of the clusters are $N_1=177$,
$N_2=130$ and $N_3=126$ particles.}
\end{figure}

We begin the discussion of our results with
Fig.~\ref{Cl_analysis}, which represents the time evolution of the
orientational order parameter $\mathcal{Q}_6(t)$, of  the number
of solidlike particles $s^{(1)}(t)$ and of the number of clusters
$s^{(0)}(t)$. These curves are computed \emph{from a single run}
in a sample sheared with the rate $\dot{\gamma}=0.005\tau^{-1}$ at
the temperature $T=0.03\epsilon/k_B$. All these quantities are
correlated and illustrate clearly the transition the initially
amorphous system into an ordered state, which takes place rapidly
after a system-dependent incubation time. Further, the transition
is also seen from the growth  in the average cluster size
presented in Fig.~\ref{correl_length}. The clear transition from
an amorphous phase to a strong crystalline order after an
incubation time, during which the system stays amorphous,
distinguishes this case from Lennard-Jonesium
\cite{Mokshin/Barrat}. Such a behaviour, in which the incubation
time is a statistical, sample-dependent quantity, is typical for a
nucleation process. After the transition, the system is
transformed essentially into a FCC structure, as  is clearly
evident from the snapshots and  from the radial distribution of
particles constitutive of  the ordered structures, $g_{cl}(r)$
(inset of Fig.~\ref{Cl_analysis}(a)). Both the cluster size
distribution and the structure of the clusters are essentially
independent of shear rate after the transition is completed,
indicating that the local structure is not shear rate dependent.
It should be noticed that ordering takes place primarily  away
from the walls, as is seen from the density profiles of solidlike
particles presented in Fig.~\ref{profile}. Similar profiles were
observed for all considered samples under shear and for all
considered strain rates and temperatures. Such a behavior points
to the homogeneous scenario of the nucleation. The evolution of
the largest nucleus as a function of time is shown in Figs.
\ref{Cl_evol} and \ref{largest_cl}.

After the transition into an ordered phase, which is accompanied
by the growth of amount of solidlike particles $s^{(1)}$ [see
Fig.~\ref{Cl_analysis}(b)], the total number of clusters $s^{(0)}$
[see Fig.~\ref{Cl_analysis}(c)] as well as the size of the largest
cluster presented in Fig.~\ref{largest_cl} become essentially
stationary.  The system achieves a high level of structural
ordering, which is, however, far from  perfect. When  $\sim 83~\%$
of the bulk becomes included into an ordered phase, no further
ordering is observed.  Remaining particles formative the
disordered ranges are located mainly near the walls. Although the
statistical properties remain essentially constant, the size of
the largest cluster displays  considerable fluctuations
(Fig.~\ref{largest_cl}). The system is made of nanocrystalline
grains containing at most a few hundred particles (see inset of
Fig.~\ref{largest_cl}), which are prevented by the shear flow from
growing further but still exchange particles under the influence
of the mechanical noise generated by the shear flow.

\begin{figure}[tbh]
\centerline{\psfig{figure=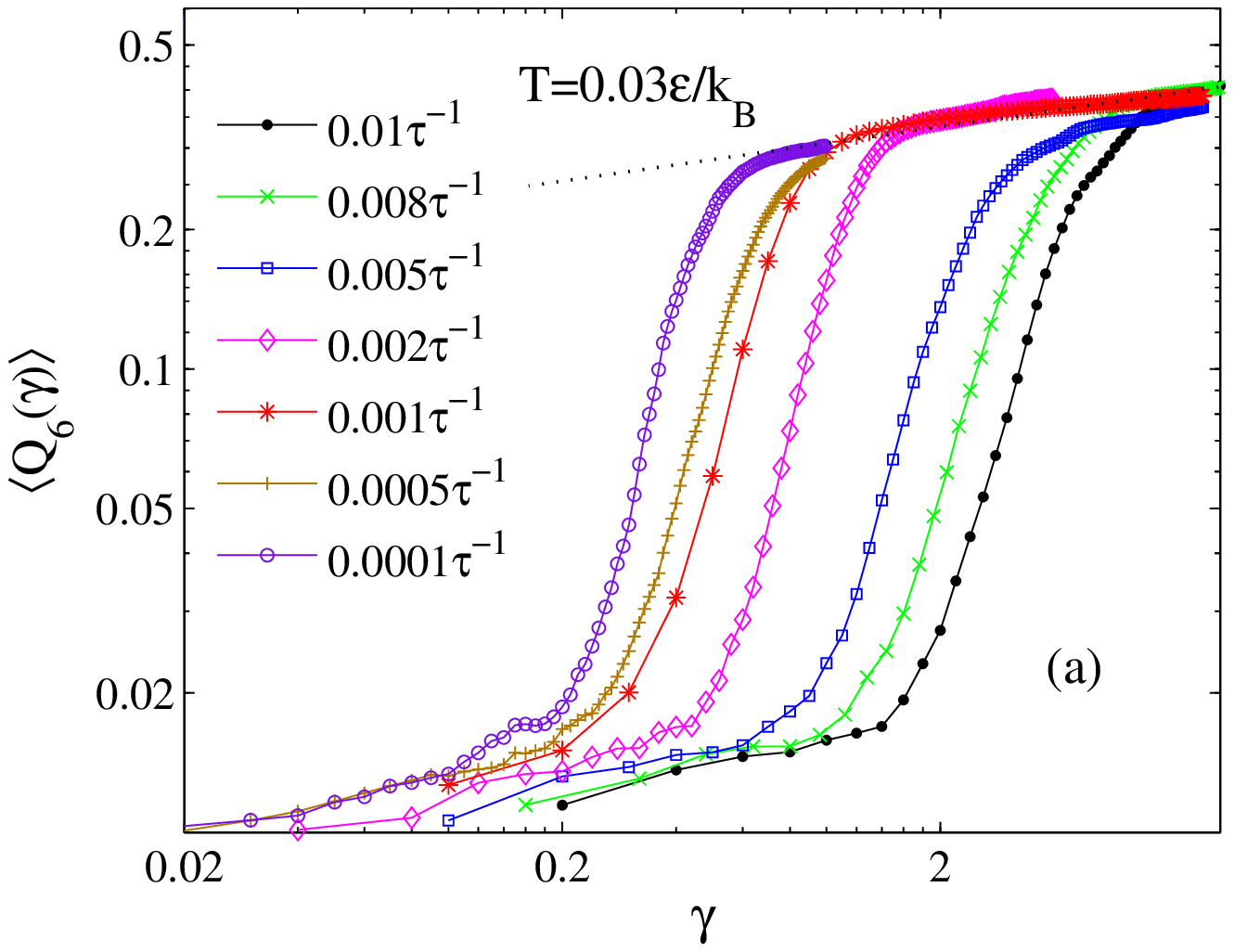,height=6.5cm,angle=0}}
\centerline{\psfig{figure=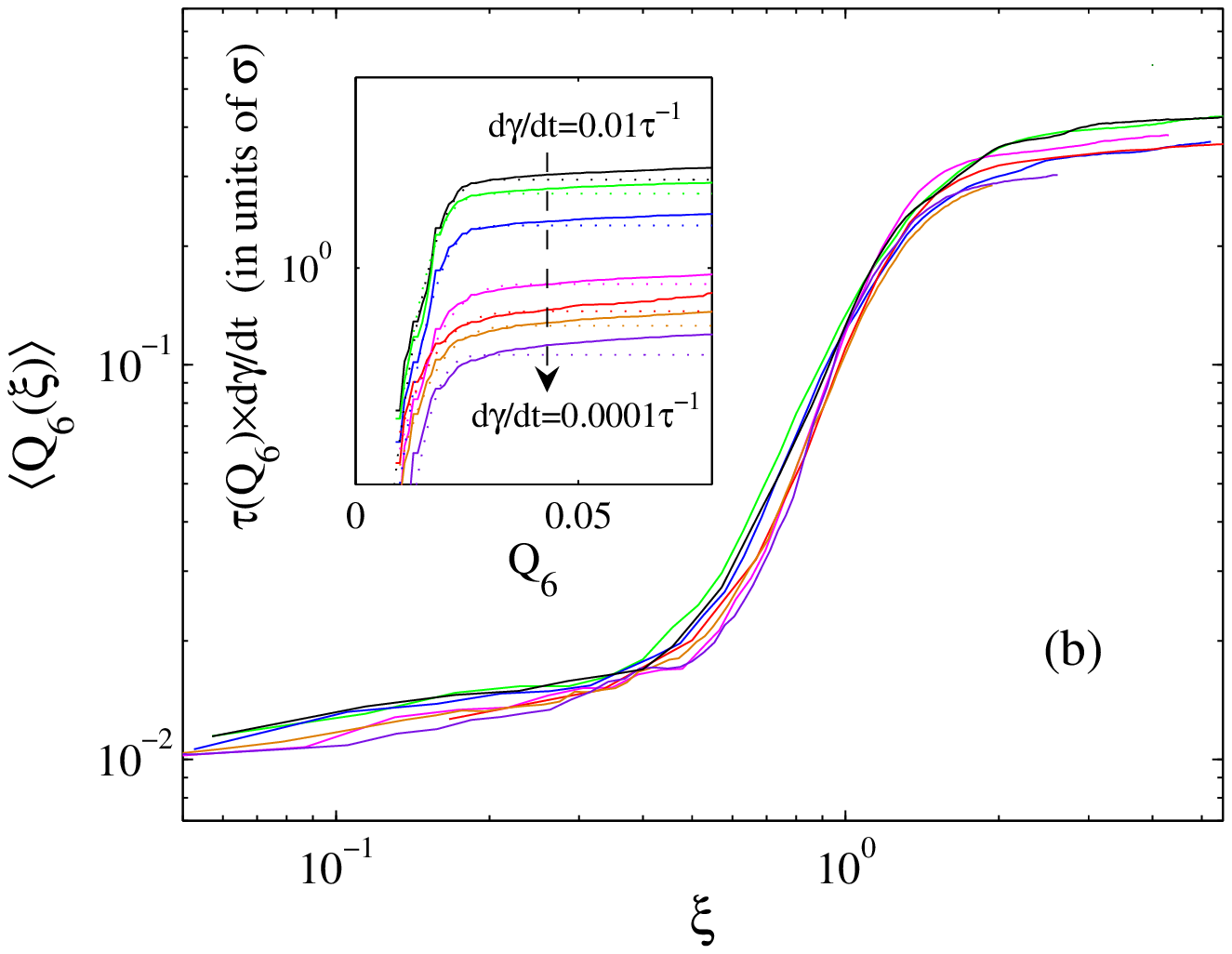,height=6.5cm,angle=0}}
\caption{\label{1} (Color online) (a) Orientational bond-order
parameter of a sheared system (amorphous initially)
\textit{versus} strain. Each curve corresponds to a  fixed strain
rate $\dot{\gamma} \in [0.0001, 0.01]$$\tau^{-1}$ and is averaged
over results obtained for at least $50$  independent samples,
which were sheared under the same conditions at a  temperature
$T=0.03\epsilon/k_B$. (b) Main: Orientational bond-order parameter
\textit{versus}  rescaled transition strains $\xi =
\gamma/\gamma_{n}$. Inset: Mean first passage time distributions
multiplied by the corresponding strain rates. The solid lines are
results of simulations, the dotted lines present the fits to
Eq.~\Ref{MFPT_distr}.}
\end{figure}
Until now we have discussed the strain history of a single,
typical sample. We now turn to the consideration of the
statistical properties of this strain history by collecting
statistics on $50$ and more independent runs. The evolution of
orientational bond-order parameter $\mathcal{Q}_6(\gamma)$ (with
$\gamma=\dot{\gamma}t$) at the temperature $T=0.03$$\epsilon/k_B$
and at various values of the strain rate $\dot{\gamma}$ is
presented in Fig.~\ref{1}(a).  These curves display a marked
transition at a well defined value $\gamma_n$  of the strain,
which increases with shear rate.  Obviously, the transition can be
characterized by a transition rate $\tau_n^{-1}$, which we
evaluate using the mean first-passage time (MFPT) method. This
method \cite{Hanggi,Wedekinda1,Wedekinda2}  focuses on the average
time $\tau$ needed for the system to reach \emph{for the first
time} some state with a defined value of the reaction coordinate
$\phi \rightarrow \phi_{1},~ \phi_{2},~ \phi_{3},~ ..., ~~ \phi\in
\mathbb{R}$, where $\mathbb{R}$ is the domain of states
\cite{MFPT_th}. As a rule, this allows one to evaluate the
distribution of time $\tau(\mathcal{\phi})$ over the reaction
coordinate $\phi$. For an activated process, the transition rate
$\tau_n^{-1}$ is simply determined by the inverse of the MFPT, at
which a plateau is reached, i.e. $\tau_n^{-1} \equiv
\tau(\mathcal{\phi}_{p})^{-1}$, whereas the total $\tau(\phi)$ can
be fitted by
\begin{equation}
\tau(\phi) = \frac{\tau_n}{2}(1+\textrm{erf}((\phi-\phi^*)c)),
\label{MFPT_distr}
\end{equation}
where $\textrm{erf}(\ldots)$ is the ordinary error function, $c$
is the local curvature around the top of the barrier and $\phi^*$
corresponds to the critical value of reaction coordinate (for
details, see Ref. \cite{Wedekinda1}).

The transition rates calculated  within the framework of the  MFPT
approach with the order parameter as the reaction coordinate,
$\mathcal{Q}_6$ $\rightarrow$ $\phi$, were used to rescale the
strain dependencies presented in Fig.~\ref{1}(a). So,
Fig.~\ref{1}(b)  shows  the  order parameter [same with
Fig.~\ref{1}(a)] with the rescaled argument $\xi =\gamma/\gamma_n=
t/\tau_n$, where $\gamma_n = \dot{\gamma}\tau_n$ is the
\emph{transition strain}. As can be seen, all curves collapse onto
a single master curve, which is indicative of a common mechanism
for the transition scenario.   The transition takes place in
 three stages, which, in turn, reflect
the influence of the external drive on the inherent dynamics of
the system. The first initial stage, where the system remains
mainly in an amorphous phase, is characterized by shear-induced
``unjamming'' of the system and by the fluidization of the glass
\cite{Barrat_Kurchan}. Due to the shear the diffusivity grows at
this stage and the potential energy increases. The
\emph{characteristic} time scale associated with inherent dynamics
is defined here by the relaxation time at a finite shear rate,
$\tau_\alpha(\dot{\gamma})$ (see Ref. \cite{Barrat_Kurchan}). The
second stage corresponds directly to the transition of the
amorphous system into an ordered state and is caused by cluster
nucleation  \cite{Mokshin/Barrat}. Its appearance can be
characterized by the transition time scale $\tau_n$ (or the
transition strain~$\gamma_n$). During  this stage the order
parameter grows rapidly, following a power law $\propto \xi^{3\pm
0.1}$.  In the third stage, shear initiates rearrangements in the
ordered structures and prevents the system from undergoing a
complete  crystallization,  the order parameter increases only
very slowly and probably saturates for large strains
\cite{Allen/Frenkel,Cerda}.

\begin{figure}[tbh]
\centerline{\psfig{figure=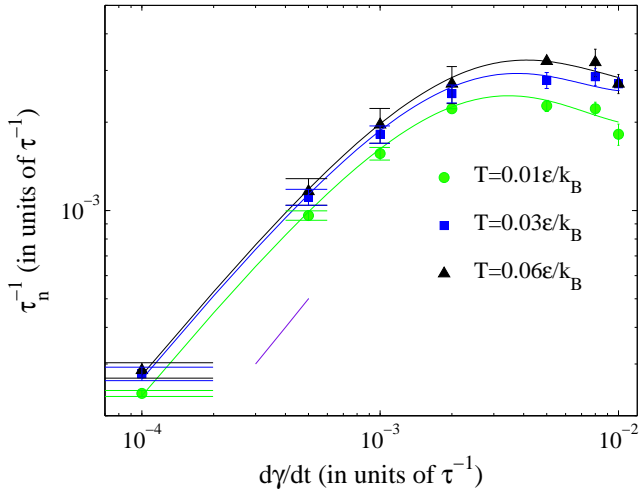,height=7cm,angle=0}}
\caption{\label{2} (Color online) Transition rate \textit{versus}
strain rate at three different temperatures. The points with error
bars are the data extracted from our molecular dynamics
simulations. The solid curves are the fits using
Eqs.~\Ref{non_eq_Ar} and \Ref{expansion_for_G}. The short straight
line corresponds to a unit slope.}
\end{figure}
The extracted values of the transition rates $\tau_n^{-1}$ at
different strain rates and three temperatures are presented in
Fig.~\ref{2}. At small values of $\dot{\gamma}$,   the transition
rates increase with strain rate in a  manner close to a power law.
Note that the rates are practically independent of temperature at
these low values of $\dot{\gamma}$. However, with the subsequent
increase of $\dot{\gamma}$ the growth of transition rates becomes
slower and temperature dependent. Eventually, starting with
$\dot{\gamma} \approx 0.005\tau^{-1}$ transition rates decrease
with the strain rate $\dot{\gamma}$.

To describe the  transition of a system into an ordered phase,
the ordinary practice is  to use the Arrhenius-like equation
\begin{equation}
\tau_{n}^{-1}=\nu_0 \exp(-\beta \Delta E),~~~ \label{Arrhenius_tr}
\end{equation}
where $\nu_0$ is the attempt frequency (or kinetic prefactor) to
overcome a barrier $\Delta E$ and $\beta = 1/(k_B T)$ is the
inverse thermal energy; $G \equiv \beta \Delta E$. Two time scales
arise naturally in this equation, $\tau_n$ and $\nu_0^{-1}$.
Identifying a rate $\tau_{n}^{-1}$ with the nucleation rate
\cite{Wolde} and the barrier $\Delta E$ with the free energy cost
to form a cluster of the critical size,  Eq. \Ref{Arrhenius_tr} is
the main equation of CNT. The use  of Eq. \Ref{Arrhenius_tr} for
systems near thermal equilibrium is straightforward: from the
measured transition rates $\tau_n^{-1}$ at  different temperatures
one can easily define the thermodynamic barrier $\Delta E$ as well
as the kinetic term $\nu_0$. However, the use of this equation for
non-equilibrium cases that violate detailed balance is not
generally justified, although there are some indications that a
similar approach using an ``effective temperature'' could be
relevant \cite{Ilg/Barrat,Haxton_Liu}.

Nucleation events in systems under shear have been identified in
Ref. \cite{Blaak2} with the conclusion that the thermodynamic
barrier $\Delta E$ is an increasing function of shear rate. We
propose an interpretation of our results along the same lines by
making a number of reasonable assumptions, namely:

\noindent (i) A formula similar to  \Ref{Arrhenius_tr}  can be
used providing an unknown effective temperature
$T_{\mathrm{eff}}$, which is introduced to replace the usual bath
temperature $T=1/(k_B \beta)$.

\noindent (ii) The kinetic prefactor is set by the time scale for
relaxation in a system under shear, which is inversely
proportional to the shear rate in this low temperature regime
\cite{Varnik_private}. Hence $\nu_0 \propto \vert \dot{ \gamma}
\vert$.

While the assumption (i) results in the generalization of the
energy barrier height $G = \beta \Delta E$ in
Eq.~\Ref{Arrhenius_tr} to the $\dot{\gamma}$-dependent term
\cite{Duff,Blaak2}, $G(\dot{\gamma},T) = \Delta E / (k_B
T_{\mathrm{eff}})$,  assumption (ii) is motivated by the fact that the
kinetic processes that can be observed in
 a glass are dominated by the  influence of the shear. As a result,
Eq.~\Ref{Arrhenius_tr} can be rewritten as
\begin{equation}
\label{non_eq_Ar} \tau_{n}^{-1} \propto \vert \dot{\gamma} \vert
\exp[-G(\dot{\gamma},T)].
\end{equation}
This is justified by  the fact that the crystal nucleation in a
glassy system free from any drive, e.g. $\dot{\gamma} \to 0$,
becomes at very deep supercooling extremely rare event, i.e.
$\tau_{n}^{-1} \to 0$, practically unobservable within the
time-scale of the experiment as  structural rearrangements
are inhibited by the very  high viscosity \cite{Schmelzer}.

These assumptions are not sufficient to determine independently
the effective temperature and the barrier $\Delta E$, both of
which, as can be expected, depend on shear rate and on the actual,
thermodynamic temperature. However, the variation of a
thermodynamic factor $G(\dot{\gamma} , T)$ with shear can be
defined. A first clear observation from Fig.~\ref{2} is that
$G(\dot{\gamma} , T)$ must be typically of order unity. If we make
the reasonable assumption that the barrier $\Delta E$ is of the
same order of magnitude as in usual supercooled liquids (typically
of order $\epsilon$, where $\epsilon$ is the energy scale of the
interaction), it appears that the usual activation at the
temperatures under study ($T<0.06$$\epsilon/k_B$) is  unable to
account for our observations. Moreover, any effective temperature
$T_{\mathrm{eff}}$, which would be used to explain the observed
values of $G(\dot{\gamma} , T)$, should depend on shear rate only
weakly.

\begin{figure}[tbh]
\centerline{\psfig{figure=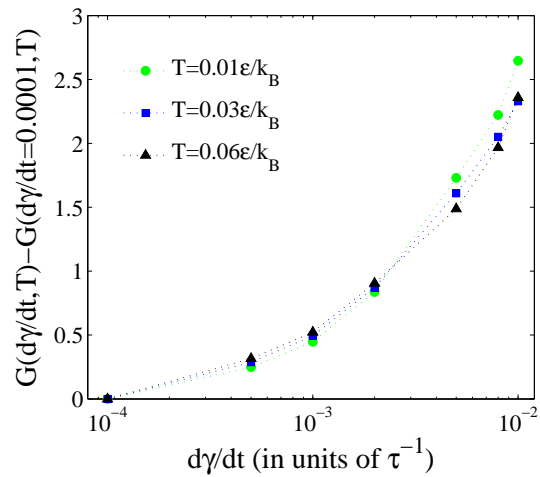,height=7cm,angle=0}}
\caption{\label{3} (Color online)  Change of the exponent of
thermodynamic factor $G(\dot{\gamma},T)$ with the strain rate in
regard to the case with $\dot{\gamma} = 0.0001\tau^{-1}$.}
\end{figure}
\begin{table}[ht]
\caption{Numerical data for different temperatures $T$ on the
parameter $\mathcal{A}$ and the correction factors $B_1$, $B_2$
and $B_3$ to the equilibrium barrier height $G^{(0)}$ as obtained
from the fitted simulation data.}
\begin{ruledtabular}
\bigskip
    \begin{tabular}{ccccc}
    $T$ ($\epsilon/k_B$) & $\mathcal{A}$ & $B_1$ ($\tau$) & $B_2$ ($\tau^2$) & $B_3$ ($\tau^3$) \\
    \hline
    0.01 & 2.44 & 430 & $-230 \times 10^2$ & $500 \times 10^3$ \\
    0.03 & 2.78 & 417 & $-229 \times 10^2$ & $505 \times 10^3$ \\
    0.06 & 2.86 & 390 & $-210 \times 10^2$ & $510 \times 10^3$ \\
    \end{tabular}
\end{ruledtabular}
\label{Tab:data}
\end{table}
The shear rate dependence of $G(\dot{\gamma} , T)$ is presented in
Fig.~\ref{3}. As the effective temperature can only increase with
shear rate, the data clearly shows that the energy barrier must
increase with shear rate. This is consistent with the findings of
Ref. \cite{Blaak2}. However, we found that the parabolic fit (in
the variable $\dot{\gamma}$) of $G(\dot{\gamma}, T)$, which was
used in Ref. \cite{Blaak2} to treat the data, is not sufficient to
describe our results. Instead we use a cubic dependence of shear
rate:
\begin{equation}
G(\dot{\gamma},T)=G^{(0)}  + \sum_{n=1}^3 B_n(T) \vert
\dot{\gamma}\vert^n, \label{expansion_for_G}
\end{equation}
where $G^{(0)}$ characterizes the limit of the energy barrier
hight at zero shear rate and $B_n(T)$ are the correction factors.
Equation~\Ref{expansion_for_G} is an expansion in powers of the
dimensionless strain rate $\dot{\gamma}\tau$ as a small parameter,
$\dot{\gamma}\tau \in [0.00001,\;0.01]$, for $G(\dot{\gamma},T)$
about the equilibrium value $G^{(0)}$. Then, Eq.~\Ref{non_eq_Ar}
can be extended to the following form:
\begin{equation}
\tau_n^{-1} = \mathcal{A} \vert \dot{\gamma} \vert \; \mathrm{exp}
\left ( -\sum_{n=1}^{3} B_n(T) \vert \dot{\gamma} \vert^n \right
), \label{extend_CNT}
\end{equation}
where $\mathcal{A} \propto \exp{[-G^{(0)}]}$ characterizes the
zero shear rate limit  of the thermodynamic  factor. The fits of
Eq.~\Ref{extend_CNT} are shown as continuous lines in Fig.~\ref{2}
and the obtained numerical values of all parameters involved in
Eq.~\Ref{extend_CNT} are given in Table~\ref{Tab:data}. Note that
Eq.~\Ref{extend_CNT} should be considered as an attempt to extend
the main equation of CNT for non-equilibrium case \cite{Blaak2} of
a glassy system under shear at very deep supercooling, where the
inherent structural rearrangements are extremely slow and the
external shear affects the transport properties directly as was
discussed in Refs.~\cite{Blaak2,Butler}. The fact that the
coefficients $B_n(T)$ are, in fact, almost independent of the
temperature shows that the process is activated by shear, rather
than by thermal fluctuations.

In order to obtain an independent determination of the effective
temperature, it would be necessary to determine an energy barrier
$E_0$ at zero temperature and zero shear rate for the formation of
a crystalline nucleus, and to obtain $T_{\textrm{eff}}$ as the
ratio $E_0/G^{(0)}$. Unfortunately no method is known to obtain
$E_0$ directly. A possibility, which we have not pursued in this
work, would be to extrapolate results from high temperatures. As a
result, our expression for the nucleation rate involving an
effective temperature cannot be seen as a direct validation of
this concept, but rather as a plausible extrapolation of other
results \cite{Haxton_Liu,Ilg/Barrat}.

Our results demonstrate that the formation of a nanocrystalline
phase in a driven, low temperature system proceeds via a
nucleation mechanism, not dissimilar to the one observed in the
standard liquids at moderate undercooling.  Nucleation theory in
its classical form is not applicable, in view of the very low
temperatures. However, our results are consistent with a
nucleation approach that would involve an effective temperature
and a nucleation barrier, which increases with shear rate, as
already reported in supercooled liquids.

This work could be extended in several directions, in particular:

(i) Metallic glasses often have  quasi-crystalline phases. For the
Dzugutov system studied here,  the quasi-crystalline phase can
coexist with the liquid (or amorphous)  and crystalline phases
(see Fig.~$4$ in Ref. \cite{Roth_PRB_2005}), and the spontaneous
growth of quasi-crystals in this system was recently investigated
by Glotzer \cite{Glotzer}.  It could be interesting to study the
competition between quasi-crystal formation and crystal nucleation
under an external drive.

(ii) Glassy materials are characterized by a waiting time
dependence of their properties, or  aging, and a state that
depends on the cooling rate. The cooling process and  the waiting
time  could have an influence on the ordering of the system  under
an  external driving.

(iii) Although the  one-component system considered in this study
is a good glass-former, bulk metallic glasses are, as a rule,
multicomponent (five different kinds of atoms, at least).
Therefore, it would be of interest to study the influence of
structural inhomogeneity appeared from polydispersity on the
induced structural ordering.

We acknowledge useful discussions with A. Tanguy, T. Biben, R.
Khusnutdinoff and R. Yulmetyev  and an interesting correspondence
with  J. Wedekind and F. Varnik. This work was supported  by ANR
project ``SLLOCDYN'' and partially by RFBR (Grant No.
08-02-00123-a).

\bibliographystyle{unsrt}

\begin{thebibliography}{99}
\bibitem{footnote1} The next reduced units are used throughout this paper. Quantities
corresponding to distances and energies are expressed in the units
$\sigma$ and $\epsilon$ of the Dzugutov potential. The reduced
time unit is $\tau = \sigma \sqrt{m_0/\epsilon}$, where $m_0$ is a
particle mass. The temperature is in units of $\epsilon/k_B$,
whereas the pressure is in units of $\epsilon/\sigma^3$.
\end{thebibliography}

\end{document}